\documentclass[]{spie}  %>>> use for US letter paper
\usepackage{amsmath,amsfonts,amssymb}
\usepackage{graphicx}
\usepackage[colorlinks=true, allcolors=blue]{hyperref}

\usepackage{subcaption}

\title{Segmentation of Multiple Myeloma Plasma Cells in Microscopy Images with Noisy Labels}

\author[a,*]{Álvaro García Faura}
\author[a,b,*]{Dejan Štepec}
\author[a,b]{Tomaž Martinčič}
\author[b]{Danijel Skočaj}
\affil[a]{XLAB d.o.o., Pot za Brdom 100, 1000, Ljubljana, Slovenia}
\affil[b]{University of Ljubljana, Faculty of Computer and Information Science, Večna pot 113, 1000 Ljubljana, Slovenia}
\affil[*]{These authors contributed equally to this work.}

\authorinfo{Further author information: (Send correspondence to A.G.F.)\\A.G.F.: E-mail: alvaro.garcia.faura@xlab.si}

\begin{document} 
\maketitle

\begin{abstract}
A key component towards an improved and fast cancer diagnosis is the development of computer-assisted tools. In this article, we present the solution that won the SegPC-2021 competition\footnote[1]{https://segpc-2021.grand-challenge.org/SegPC-2021/} for the segmentation of multiple myeloma plasma cells in microscopy images. The labels used in the competition dataset were generated semi-automatically and presented noise. To deal with it, a heavy image augmentation procedure was carried out and predictions from several models were combined using a custom ensemble strategy. State-of-the-art feature extractors and instance segmentation architectures were used, resulting in a mean Intersection-over-Union of 0.9389 on the SegPC-2021 final test set.
\end{abstract}


\section{INTRODUCTION}
\label{sec:intro}  % \label{} allows reference to this section
Computer-assisted diagnostic tools are becoming an important asset for a more effective cancer diagnosis. Such tools typically require capturing images, stain color normalization, segmentation of cells of interest, and classification to count malignant versus healthy cells. Despite the recent surge of self-supervised and unsupervised methods, the use of supervised approaches is still predominant for medical images. The main drawback of supervised methods is the need for quality labels, which are especially hard to obtain for medical image datasets because of the need for domain experts. Nevertheless, the compilation of labeled datasets will still be needed to foster advancements in specific fields, at least for evaluation purposes. The development of methods that can deal with noise and ambiguity in labels is thus crucial in order to leverage datasets generated with limited resources.

The development of automatic methods for cell segmentation in microscopy images has been studied for decades \cite{meijering2012cell}. However, it has been in recent years that the proliferation of deep learning has led to state-of-the-art results in, fairly, any field within medical image analysis \cite{litjens2017survey}. Specifically regarding segmentation, U-Net \cite{ronneberger2015u} represents a fully convolutional architecture that outperformed existing methods in cell segmentation and tracking challenges. This architecture served as a stem for many others and was recently presented in a self-configuring manner~\cite{nnunet}, which greatly simplifies hyperparameter tunning across different data modalities and surpassed highly specialized solutions on 23 public medical datasets. Similarly, Mask R-CNN~\cite{he2017mask} represents a universal approach for instance segmentation that was adapted in different domains, including cell nucleus segmentation~\cite{johnson2019automatic}.

Multiple Myeloma (MM) is a type of blood cancer, specifically, a plasma cell cancer. The first stage to build an automated diagnostic tool for MM is the robust segmentation of cells. This was the overall goal of the SegPC-2021 challenge, whose organizers provided images captured from the bone marrow aspirate slides of MM patients. The problem of segmenting plasma cells in these images is complex due to the diversity of situations that may occur. For instance, cells may appear in clusters or isolated, with varying size of their nucleus and cytoplasm, some of these touching each other and being hard to differentiate. Furthermore, the staining process is not perfect and there may be unstained cells or cells for which the cytoplasm's color resembles that of the background. The problem is challenging \textit{per se}, but the use of computer-generated imperfect labels adds complexity to it because of the associated noise and inconsistencies.

In this paper, we present a method for instance segmentation of MM plasma cells in microscopy images that were labeled using a semi-automatic procedure resulting in noisy annotations. Many works have proposed ad-hoc solutions to deal with such labels \cite{karimi2020deep}. In our work, we rely on image augmentation and model ensembles to cope with SegPC-2021 imperfect labels, which resulted in arguably better segmentation results than the actual ground-truth (GT) labels and, eventually, in the winning entry for SegPC-2021 challenge.

\section{SegPC-2021 CHALLENGE DATASET}
\label{sec:dataset}

\begin{figure}[!htb]
     \centering
     \begin{subfigure}[b]{0.24\textwidth}
         \centering
         \includegraphics[width=\textwidth]{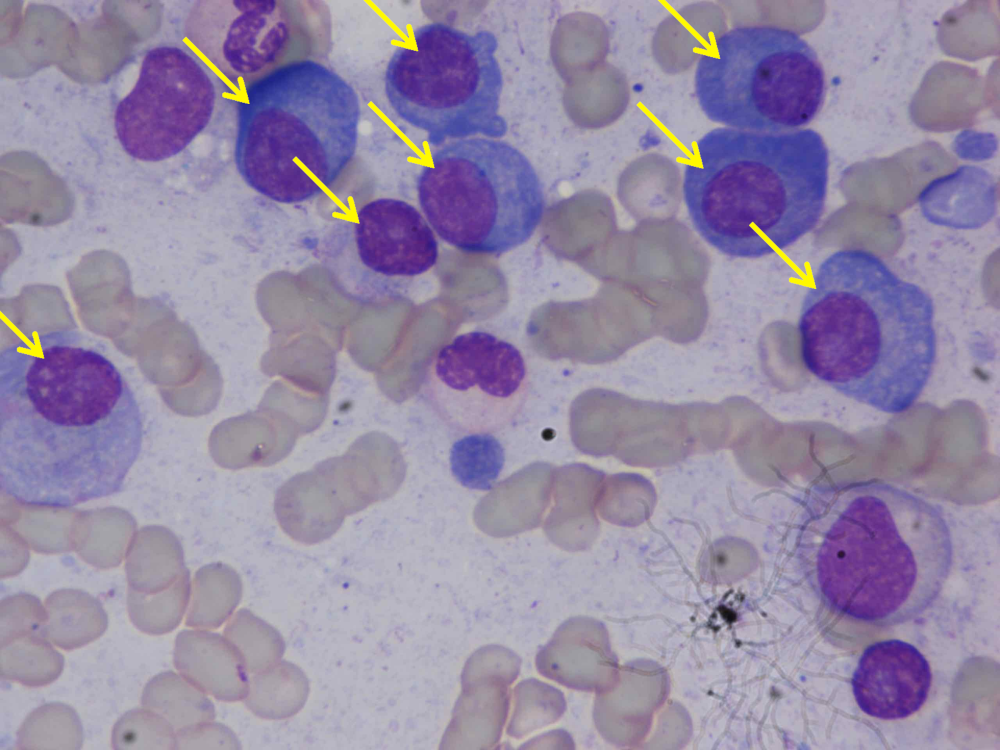}
     \end{subfigure}
          \begin{subfigure}[b]{0.24\textwidth}
         \centering
         \includegraphics[width=\textwidth]{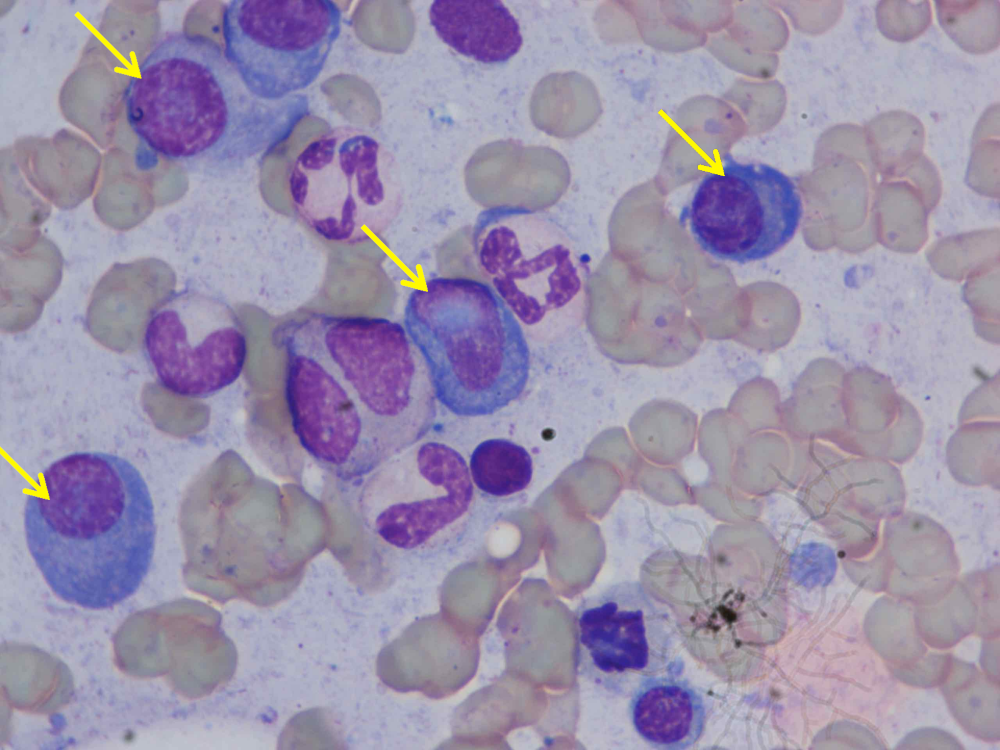}
     \end{subfigure}
     \begin{subfigure}[b]{0.24\textwidth}
         \centering
         \includegraphics[width=\textwidth]{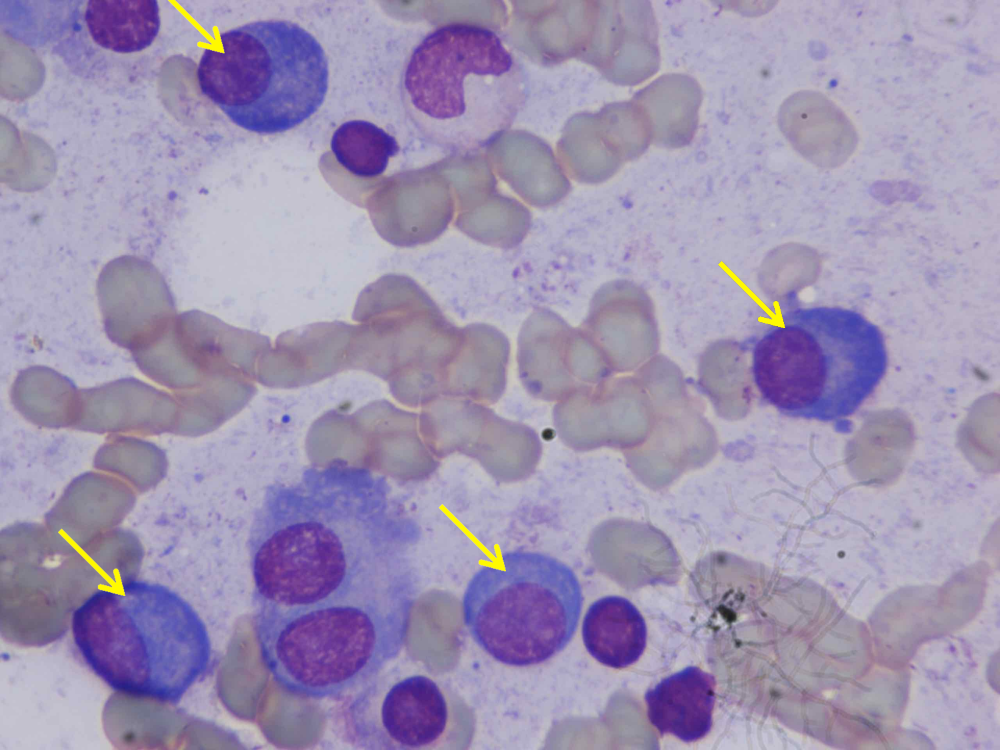}
     \end{subfigure}
     \begin{subfigure}[b]{0.24\textwidth}
         \centering
         \includegraphics[width=\textwidth]{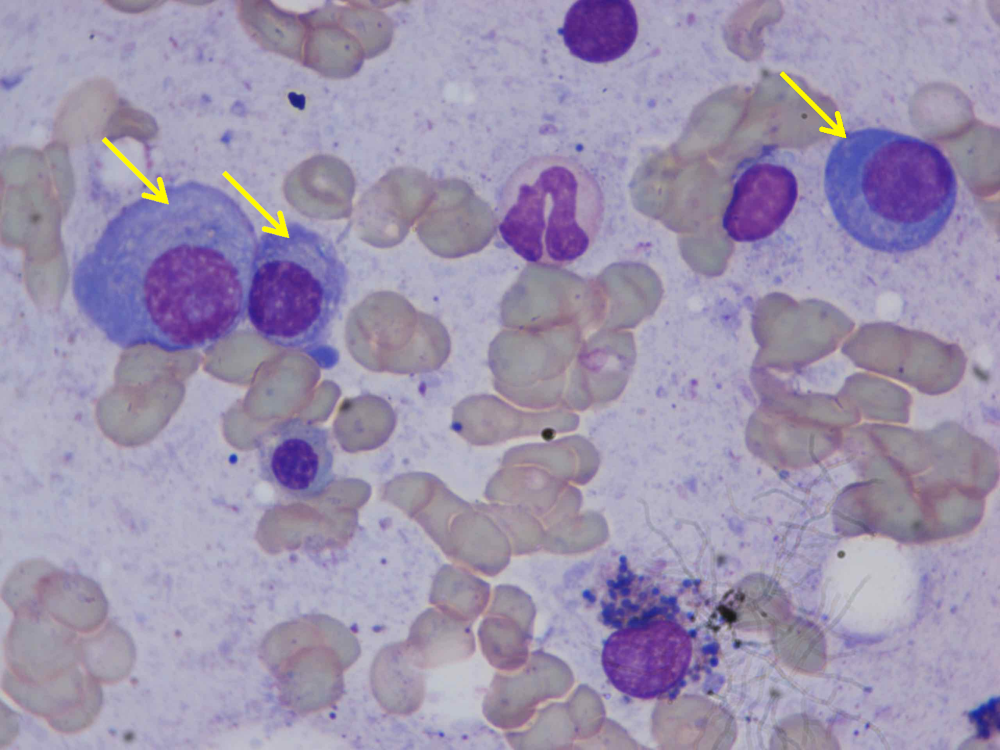}
     \end{subfigure}
     \\
     \begin{subfigure}[b]{0.24\textwidth}
         \centering
         \includegraphics[width=\textwidth]{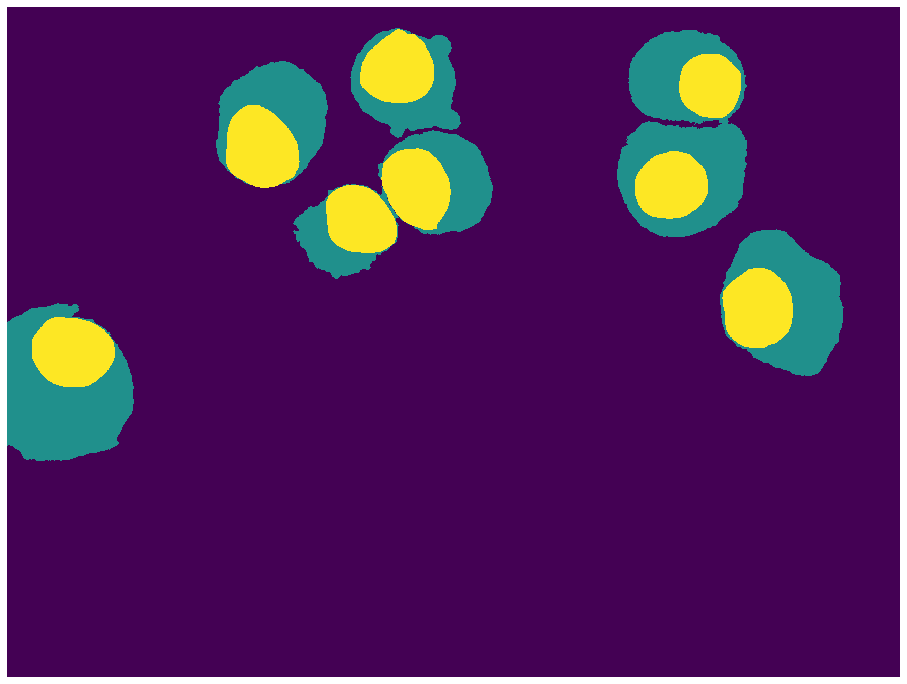}
     \end{subfigure}
      \begin{subfigure}[b]{0.24\textwidth}
         \centering
         \includegraphics[width=\textwidth]{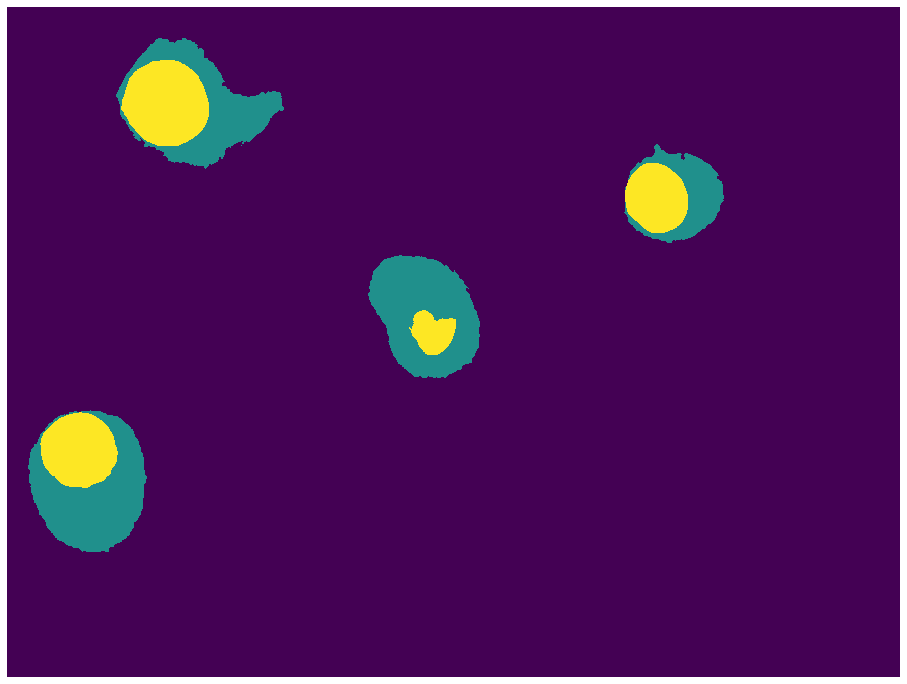}
     \end{subfigure}
     \begin{subfigure}[b]{0.24\textwidth}
         \centering
         \includegraphics[width=\textwidth]{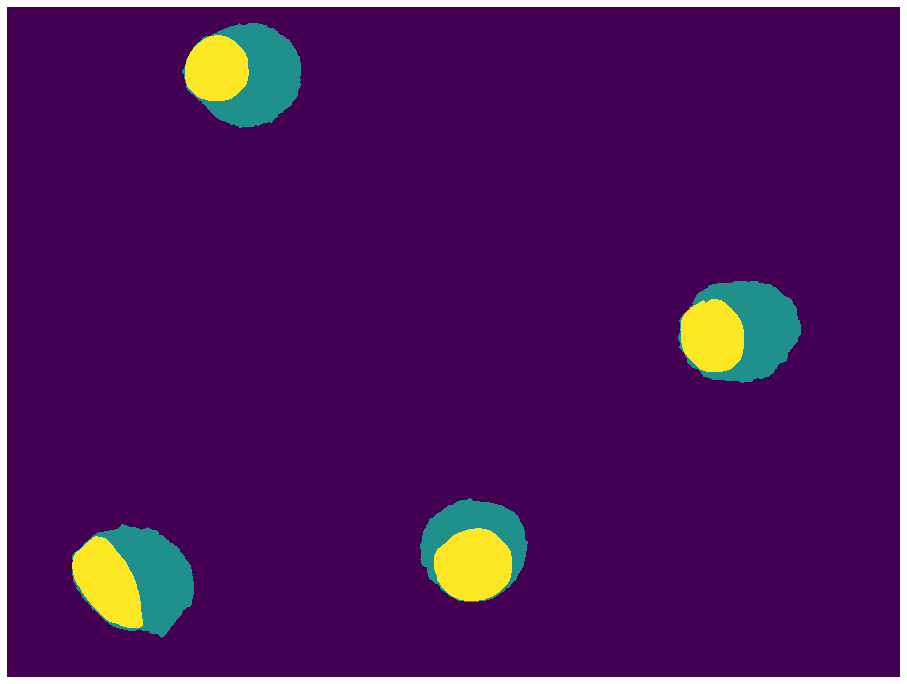}
     \end{subfigure}
     \begin{subfigure}[b]{0.24\textwidth}
         \centering
         \includegraphics[width=\textwidth]{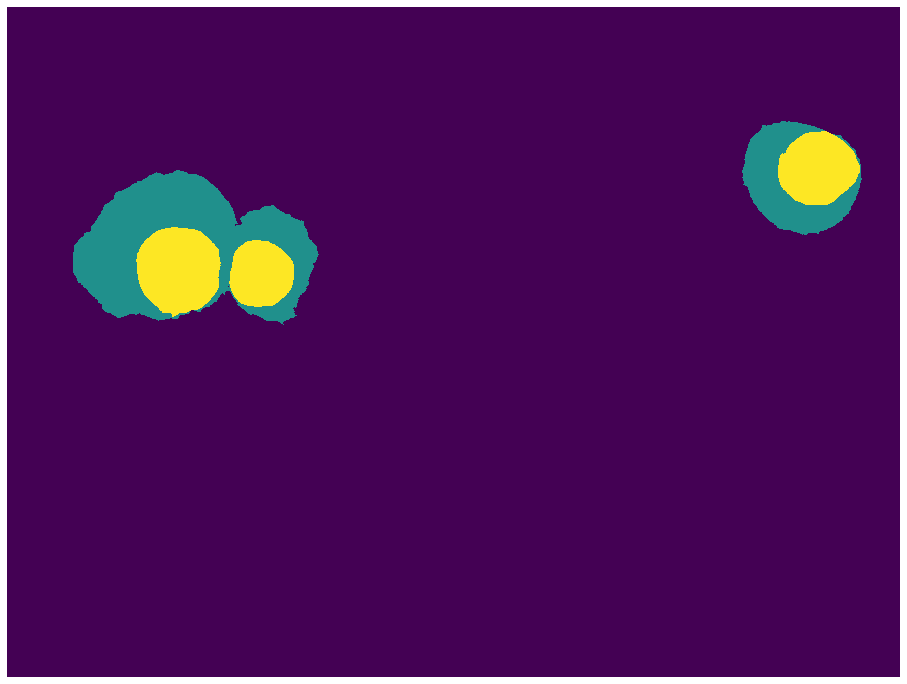}
     \end{subfigure}
        \caption{First row: weak labels provided by an expert. Cells of interest are marked with a yellow arrow (black in the original labels, changed for visibility). Second row: SegPC-2021 competition GT masks generated semi-automatically. Note that all the instances in an image are shown here together for visualization purposes, but masks are provided for each instance separately.}
        \label{fig:arrow_labels}
\end{figure}

The SegPC-2021 dataset \cite{7np1-2q42-21} consisted of images captured from slides of bone marrow aspirates collected from patients suffering from MM. The competition organizers used stain color normalization \cite{gupta2020gcti} to preprocess the images. Images were annotated in two stages. First, the cells of interest were identified and marked by an expert oncologist. Then, the nuclei of those cells were segmented using the method described in  \cite{gehlot2020ednfc}. The expert annotations were used to automatically select the relevant segmented nuclei and discard the rest, and their cytoplasm was segmented using MATLAB's Local Graph Cut tool \cite{rother2004grabcut}. This tool automatically produces segmentation masks for a region of interest that can be later refined by the user. Even though the generated instance masks corresponded only to the cells of interest identified by an oncologist, the quality of those masks was not evaluated. Some examples of original images labeled by the expert and the ground truth masks provided for the competition are included in Figure~\ref{fig:arrow_labels}.

\begin{figure}[!htb]
    \centering
    \includegraphics[width=\textwidth]{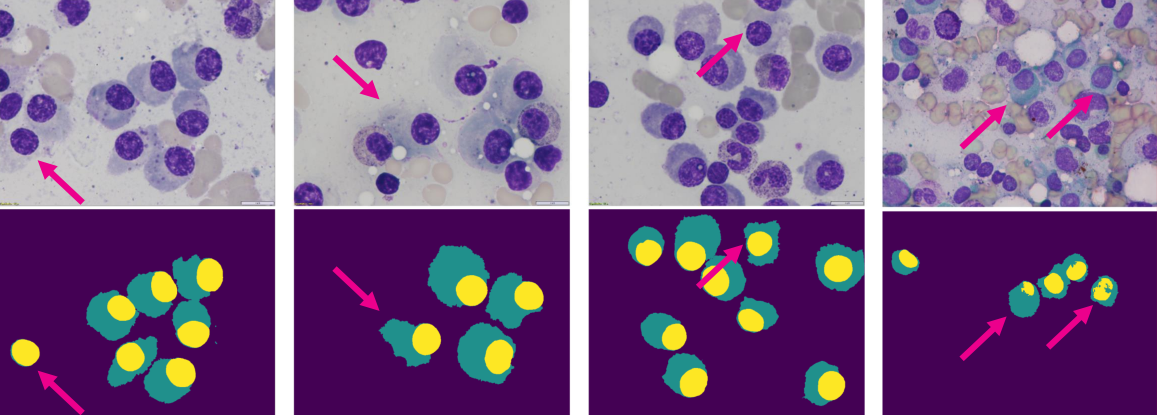}
    \caption{Examples of different types of noise in labels. First row: original images. Second row: GT labels for all the annotated cell instances in an image. Arrows are included to highlight some instances for which the imperfect labeling is clear.}
    \label{fig:noisy_labels}
\end{figure}

The use of the aforementioned semi-automated procedure to label the images in the competition dataset led to noisy and inconsistent labels. Examples of this can be seen in Figure~\ref{fig:noisy_labels}. Overall, ground truth masks are characterized by spiky contours that do not reflect the true aspect of the cell. In some cases, cytoplasm regions whose color is close to that of the background are partially and sometimes completely omitted. Although less common, nucleus masks also include noise in some images (e.g. last column in Fig.~\ref{fig:noisy_labels}). These problems require a robust method that can learn to segment cells consistently, despite the high variability seen in their morphological and color characteristics.

The competition training data for the final phase consisted of 498 images, which were additionally split into train/validation sets (20\% validation set). The final competition test set, on which all of the competitors were evaluated, consisted of 277 images. The only data used in our experiments was provided as part of the SegPC-2021 competition. No additional external data was used.

\section{METHODOLOGY}
\label{sec:methodology}
Our approach is based on instance segmentation methods. The use of semantic segmentation models would require additional post-processing to split the resulting image-level masks into different instances. We decided to avoid this and tackle the problem directly as an instance segmentation problem for simplicity.

\subsection{Instance Segmentation}
Three existing instance segmentation models have been used: Mask R-CNN \cite{he2017mask}, Hybrid Task Cascade (HTC) \cite{chen2019hybrid}, and SCNet \cite{vu2020scnet}. An in-depth analysis of these architectures is out of the scope of this article, so we kindly refer the readers to the corresponding papers. Nonetheless, as a high-level explanation, note that Mask R-CNN serves as a baseline approach both for HTC and SCNet. HTC combines Mask R-CNN and Cascade R-CNN \cite{cai2018cascade} in a novel way by interweaving the detection and segmentation tasks and including an additional semantic branch to provide spatial context. Based on HTC, the recent SCNet combines all the mask prediction stages in a single stage and moves it to the end to improve sample consistency between training and inference. Additionally, it includes a custom feature relay stage and a global context branch. All the aforementioned methods require a convolutional backbone to act as a feature extractor. In our work, we have explored the use of ResNet \cite{he2016deep} and the more recent ResNeSt \cite{zhang2020resnest}.

In the SegPC-2021 challenge, pixels belong to one of two classes, apart from the background: nucleus or cytoplasm. Unlike other classical instance segmentation datasets that include many different types of scenes, in this case, all images depict the same type of content and the two classes are related spatially. In general, a cell nucleus appears surrounded by cytoplasm, composing a whole cell.

When carrying out instance segmentation, results for different classes are given separately. This poses the problem of accurately pairing corresponding nucleus and cytoplasm instances. To ease this process, we decided to add a third class that comprises the whole cell, both nucleus and cytoplasm. This class is taken as the main segmentation result and is later combined with the other two to produce the final two-class result. The semantic branch of both HTC and SCNet models, whose purpose is to provide global context, was fed during training with binary masks for this whole-cell class including all the instances in an image.

\subsection{Image Augmentation}
Due to the limited amount of training data, heavy image augmentation has been carried out. We generated 50 additional augmented images for each image in the training set, adding up to 20298 training images. Several different kinds of augmentations were combined: geometric (scale, flip, rotate, elastic transform, piece-wise affine transform, perspective transform), color (brightness, color to gray, modification of hue and saturation), contrast (CLAHE, gamma contrast, linear contrast), blur (Gaussian, median, motion), convolutional (sharpen, emboss), and gaussian noise corruption. Augmentations were generated using the \textit{imgaug} library\footnote{https://github.com/aleju/imgaug}. Test-time augmentation was also used but limited to random flip (vertical, horizontal, and diagonal).

\subsection{Model Ensemble}
In order to provide more robust segmentation masks, a custom strategy to combine the results of different models was designed taking into account the particularities of the dataset and evaluation method of the SegPC-2021 challenge, which are described in Sections~\ref{sec:dataset} and \ref{subsec:evaluation}, respectively.

Out of a set of several trained models, one of them is selected as a reference model. To refine its predictions, majority voting is carried out for each one of the cells it segmented using one prediction per each of the remaining models. For each of these models, the instance with the largest IoU with that of the reference model is selected. A minimum IoU is needed for a prediction to participate in the voting. All the predictions that were not used in the majority voting procedure were included as part of the submission to the competition as well.

\section{EXPERIMENTS AND RESULTS}
\label{sec:exp_and_results}

\subsection{Evaluation}
\label{subsec:evaluation}
The metric used for the evaluation of the models in the SegPC-2021 is the mean Intersection-over-Union (mIoU). Specifically, for each one of the ground truth cell instances in an image, the candidate instance with the highest IoU from all the predictions for that image is selected. The mIoU is the result of accumulating the IoU of all the selected predictions and dividing it by the total number of ground truth instances. The predicted cells that are not selected as the best match for any of the ground truth cells do not penalize in any way the final score. 

\subsection{Results}

\begin{figure}[!htb]
     \centering
     \makebox[20pt]{\raisebox{22pt}{\rotatebox[origin=c]{90}{Original}}}
     \begin{subfigure}[b]{0.15\textwidth}
         \centering
         \includegraphics[width=\textwidth]{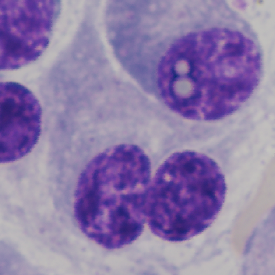}
     \end{subfigure}
      \begin{subfigure}[b]{0.15\textwidth}
         \centering
         \includegraphics[width=\textwidth]{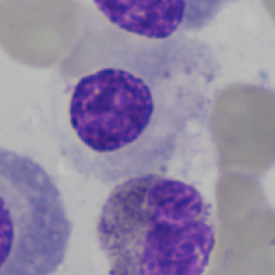}
     \end{subfigure}
     \begin{subfigure}[b]{0.15\textwidth}
         \centering
         \includegraphics[width=\textwidth]{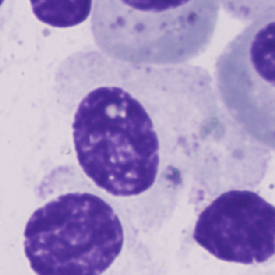}
     \end{subfigure}
     \begin{subfigure}[b]{0.15\textwidth}
         \centering
         \includegraphics[width=\textwidth]{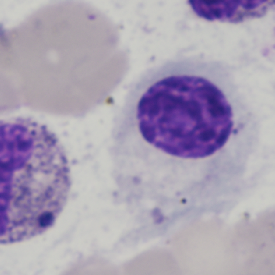}
     \end{subfigure}
     \\
     \makebox[20pt]{\raisebox{22pt}{\rotatebox[origin=c]{90}{GT}}}
     \begin{subfigure}[b]{0.15\textwidth}
         \centering
         \includegraphics[width=\textwidth]{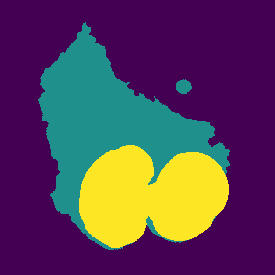}
     \end{subfigure}
     \begin{subfigure}[b]{0.15\textwidth}
         \centering
         \includegraphics[width=\textwidth]{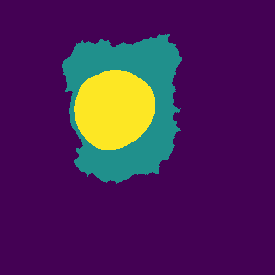}
     \end{subfigure}
     \begin{subfigure}[b]{0.15\textwidth}
         \centering
         \includegraphics[width=\textwidth]{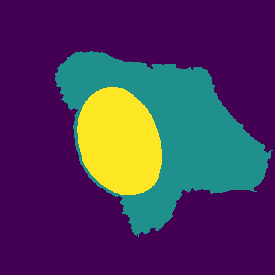}
     \end{subfigure}
     \begin{subfigure}[b]{0.15\textwidth}
         \centering
         \includegraphics[width=\textwidth]{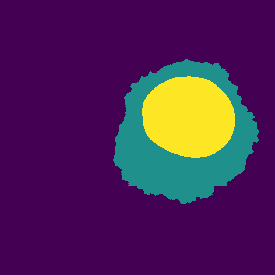}
     \end{subfigure}
     \\
     \makebox[20pt]{\raisebox{22pt}{\rotatebox[origin=c]{90}{Prediction}}}
     \begin{subfigure}[b]{0.15\textwidth}
         \centering
         \includegraphics[width=\textwidth]{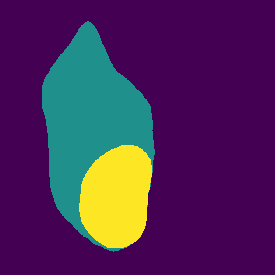}
     \end{subfigure}
     \begin{subfigure}[b]{0.15\textwidth}
         \centering
         \includegraphics[width=\textwidth]{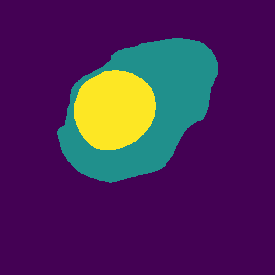}
     \end{subfigure}
     \begin{subfigure}[b]{0.15\textwidth}
         \centering
         \includegraphics[width=\textwidth]{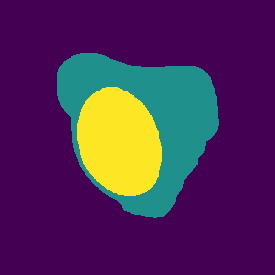}
     \end{subfigure}
     \begin{subfigure}[b]{0.15\textwidth}
         \centering
         \includegraphics[width=\textwidth]{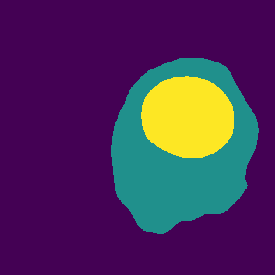}
     \end{subfigure}  
     \caption{Worst predictions done by our final model ensemble for our validation set. Images have been cropped to the cell of interest.}
     \label{fig:results_instances}
\end{figure}

To illustrate the performance of our method when dealing with noisy GT labels, we include in Figure~\ref{fig:results_instances} the four worst predictions done by the final model ensemble for our validation set. Note that the predictions correspond to the best-found match (in terms of IoU) with the GT label. In all four cases, our predictions can be argued to better adjust to the cells than the provided GT masks. In the first example, for which our best match only got an IoU of 0.6376 with the GT, we see that the issue is a mislabeling of two nuclei touching each other as a single cell. In the second example, with an IoU of 0.6752, the GT label misses part of the cytoplasm, while for the third example, in which it is hard to tell if the labeled cytoplasm corresponds entirely to that nucleus, our prediction achieves a 0.7606 IoU. Finally, in the last example the GT seems to leave out part of the cytoplasm on the bottom part, while our prediction (IoU = 0.7734) captures it adequately. 

\begin{figure}[!htb]
     \centering
     \begin{subfigure}[b]{0.25\textwidth}
         \centering
         \includegraphics[width=\textwidth]{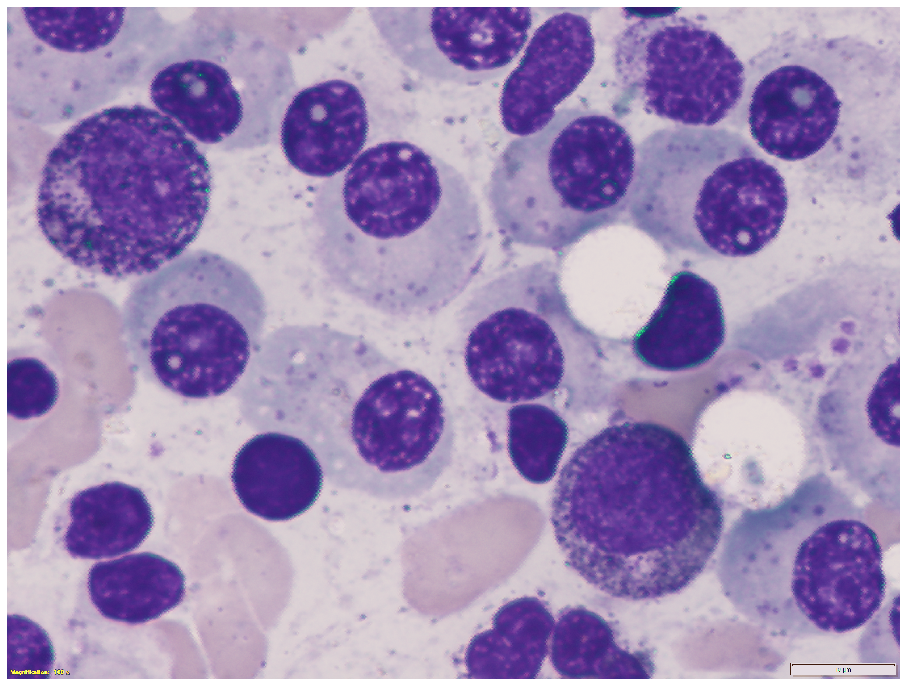}
     \end{subfigure}
     \begin{subfigure}[b]{0.25\textwidth}
         \centering
         \includegraphics[width=\textwidth]{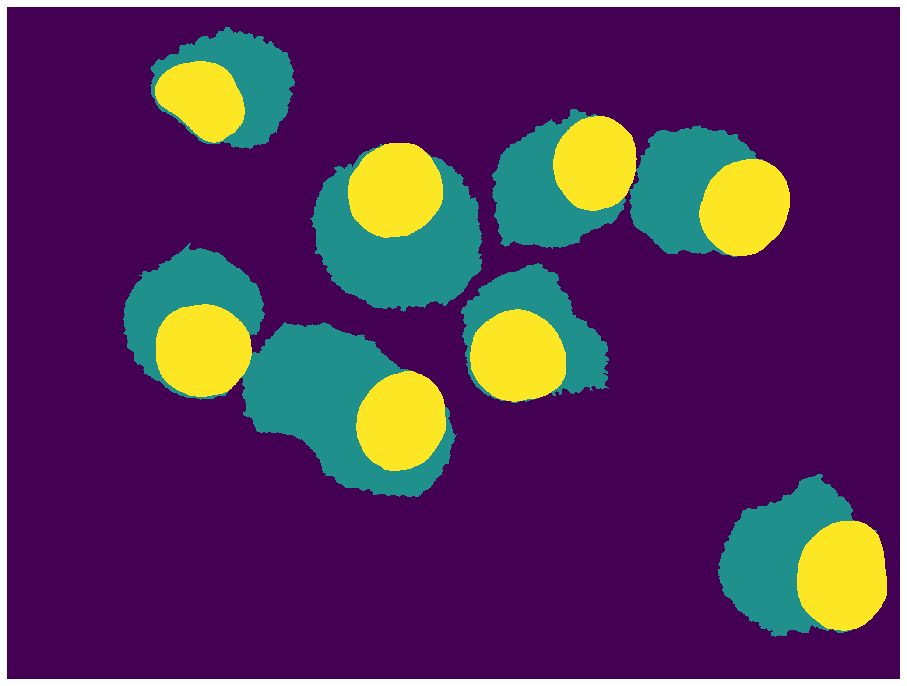}
     \end{subfigure}
     \begin{subfigure}[b]{0.25\textwidth}
         \centering
         \includegraphics[width=\textwidth]{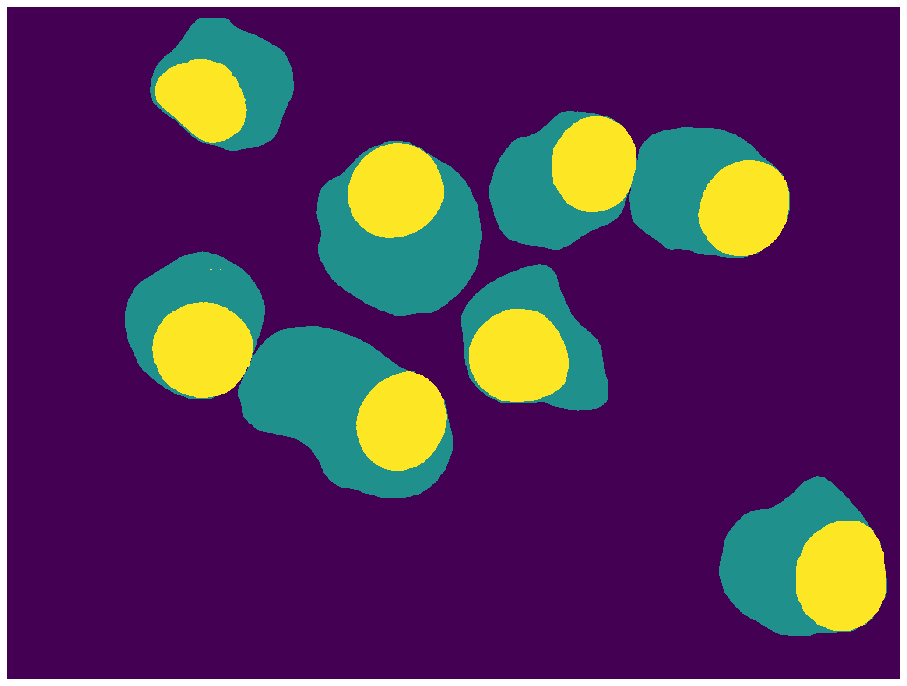}
     \end{subfigure}
     \\
     \begin{subfigure}[b]{0.25\textwidth}
         \centering
         \includegraphics[width=\textwidth]{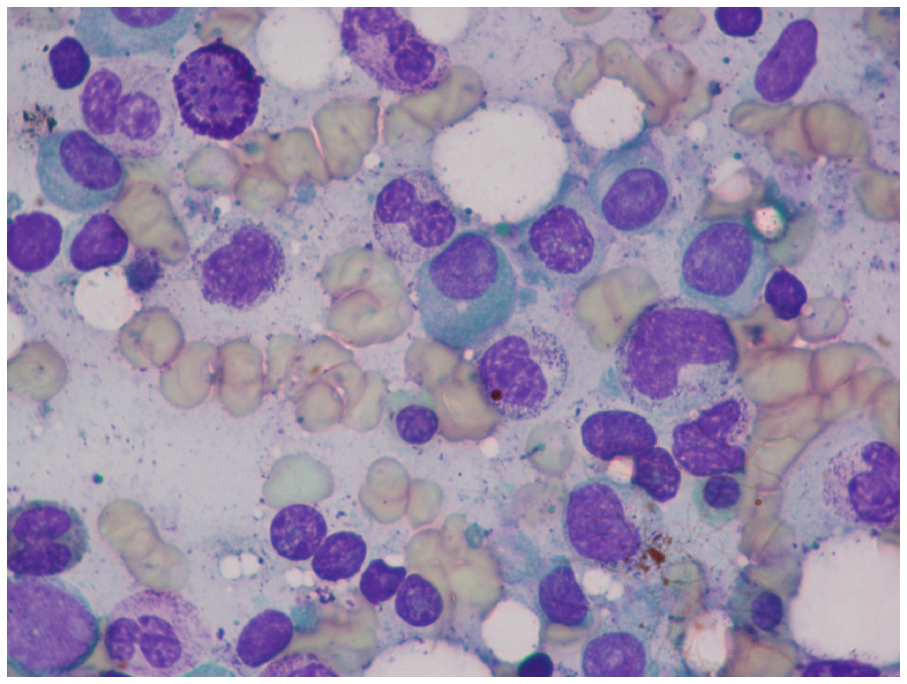}
         \caption{Original}
     \end{subfigure}
     \begin{subfigure}[b]{0.25\textwidth}
         \centering
         \includegraphics[width=\textwidth]{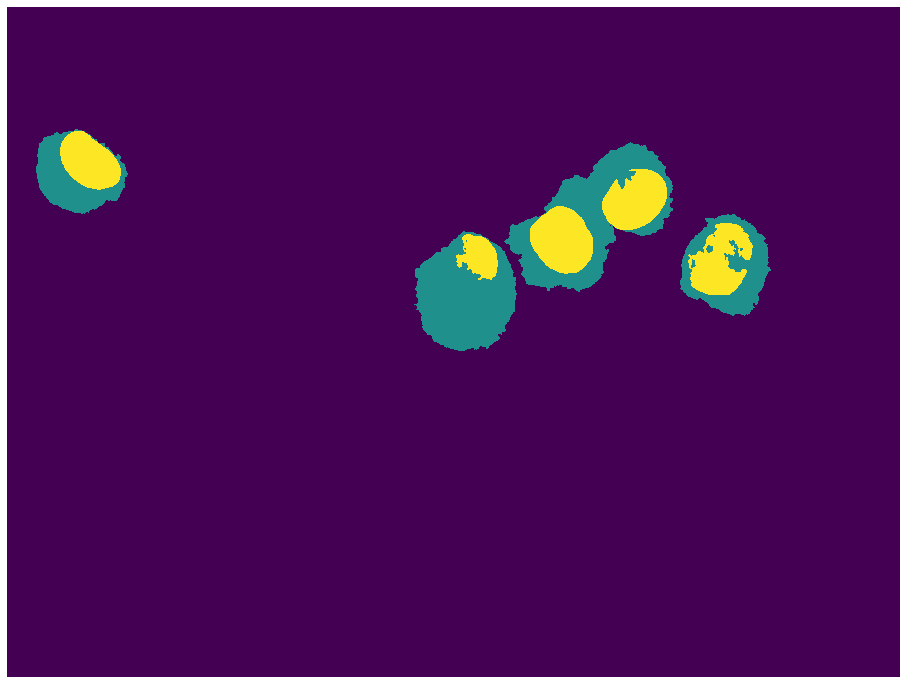}
          \caption{Ground truth}
     \end{subfigure}
     \begin{subfigure}[b]{0.25\textwidth}
         \centering
         \includegraphics[width=\textwidth]{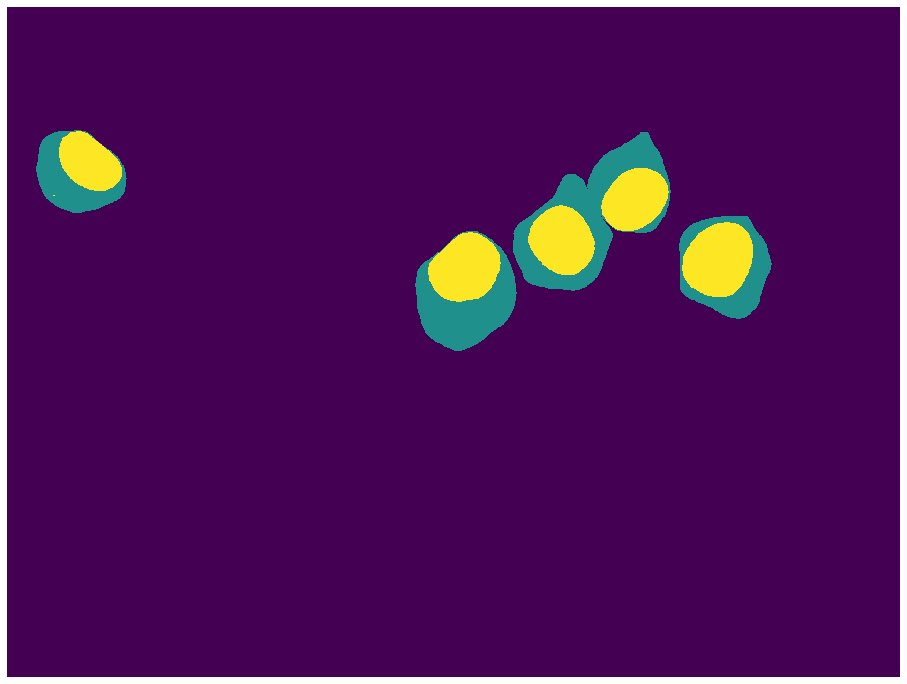}
          \caption{Prediction}
     \end{subfigure}   
        \caption{Prediction results for our final model ensemble.}
        \label{fig:results}
\end{figure}

Figure~\ref{fig:results} shows the predictions of our final ensemble for two sample images. Predictions with the highest IoU with respect to the ground truth instance masks were selected, similarly to what is done by the competition evaluation script. Therefore, not all our predictions for an image are included here. In the first row, we observe that all cells of interest were detected very accurately. It is worth noting that the spikier contours of the ground truth labels were not replicated into the predicted masks. Overall, predicted cells present softer contours which, in our opinion, reflect better the actual shape of the cells as perceived from the original images. The second example shows that our model succeeded to classify noisy nucleus labels to better match those in the original image. 

In Table~\ref{table:results}, we include the mean IoU, computed as described in Section~\ref{subsec:evaluation} for the models we trained for the SegPC-2021 competition. The best performing epoch in terms of mean Average Precision (mAP) on our validation set was selected for each of the models. This was a key aspect since considerable variations could be observed along the training process, most likely indicating over-fitting to the limited training data.

\begin{table}[htb!]
\centering
\caption{mIoU for the models trained for the SegPC-2021 competition. The models providing the 7 best results for the competition final test set (in bold) were selected to form the final ensemble.}
\begin{tabular}{ |l||c|c|c|c||c|c|c|c|}
    \hline
    Dataset & \multicolumn{4}{|c||}{Our validation set} & \multicolumn{4}{|c|}{Competition final test set} \\
    \hline
    Backbone & \multicolumn{2}{|c|}{ResNet} & \multicolumn{2}{|c||}{ResNeSt} & \multicolumn{2}{|c|}{ResNet} & \multicolumn{2}{|c|}{ResNeSt} \\
    \hline
    Backbone depth & 50 & 101 & 50 & 101 & 50 & 101 & 50 & 101 \\
    \hline \hline
    Mask R-CNN & 0.9032 & 0.9091 & 0.9154 & 0.9165 & 0.9028 & 0.9120 & 0.9162 & 0.9151 \\
    HTC & 0.9295 & 0.9280 & 0.9295 & 0.9300 & \textbf{0.9301} & \textbf{0.9285} & 0.9262 & \textbf{0.9280} \\ 
    SCNet & 0.9349 & 0.9313 & 0.9339 & 0.9324 & \textbf{0.9335} & \textbf{0.9309} & \textbf{0.9313} & \textbf{0.9314} \\
    \hline \hline
    Ensemble & \multicolumn{4}{|c||}{0.9404} & \multicolumn{4}{|c|}{\textbf{0.9389}} \\ 
    \hline
\end{tabular}
\label{table:results}
\end{table}

For any of the backbones we used, the Hybrid Task Cascade (HTC) \cite{chen2019hybrid} architecture consistently outperformed Mask R-CNN \cite{he2017mask}, and SCNet \cite{vu2020scnet} does so for both the other architectures. The best performing single model was SCNet with ResNet-50 backbone. When comparing different backbones depths, we observe that deeper ones perform better for the simpler Mask R-CNN but, overall, HTC and SCNet generate better predictions when using shallower 50-layer backbones.

\section{CONCLUSION}
\label{sec:conclusion}
In this article, we have presented the winning solution for the SegPC-2021 competition in detail. The semi-automatic labeling method used produced valuable but imperfect labels. Relying on heavy data augmentation and a custom procedure to aggregate results from different models based on majority voting, robust instance segmentation results were generated. The combination of state-of-the-art convolutional backbones and instance segmentation architectures led to a variety of single models that already showed remarkable performance. The predictions of seven of these models were aggregated and the resulting model ensemble outperformed the solutions of the rest of the participants with a mIoU of 0.9389.

\acknowledgments % equivalent to \section*{ACKNOWLEDGMENTS}       
This work was partially supported by the European Commission through the Horizon 2020 research and innovation program under grant 826121 (iPC).

% References
\bibliography{report} % bibliography data in report.bib
\bibliographystyle{spiebib} % makes bibtex use spiebib.bst

\end{document}